\documentclass[12pt]{iopart}
\usepackage[]{epsfig}
\newcommand{\be}{\begin{eqnarray}}
\newcommand{\ee}{\end{eqnarray}}

\def\theequation{\thesection.\arabic{equation}}
\begin{document}
\title[Classical dynamics and stability of collapsing thick shells of matter]
{Classical dynamics and stability of collapsing thick shells of matter}
\author{G.L.~Alberghi$^{1,2}$\footnote{\tt alberghi@bo.infn.it},
R.~Casadio$^1$\footnote{\tt casadio@bo.infn.it},
and D.~Fazi$^1$\footnote{\tt fazi@bo.infn.it}}
\address{$^1$ Dipartimento di Fisica, Universit\`a di Bologna,
and I.N.F.N, Sezione di Bologna,
Via Irnerio 46, 40126 Bologna, Italy.}
\address{$^2$ Dipartimento di Astronomia, Universit\`a di Bologna,
Via Ranzani 1, 40127 Bologna, Italy.}
%
%
%
\begin{abstract}
We study the collapse towards the gravitational radius of a macroscopic
spherical thick shell surrounding an inner massive core.
This overall electrically neutral \emph{macroshell\/} is composed by many nested
$\delta$-like massive \emph{microshells\/} which can bear non-zero electric charge,
and a possibly non-zero cosmological constant is also included.
The dynamics of the shells is described by means of Israel's (Lanczos)
junction conditions for singular hypersurfaces and, adopting a \emph{Hartree\/}
(mean field) approach, an effective Hamiltonian for the motion of each microshell
is derived which allows to check the stability of the matter composing
the macroshell.
We end by briefly commenting on the quantum effects which may arise from
the extension of our classical treatment to the semiclassical level.
\end{abstract}
\raggedbottom
\setcounter{page}{1}
\section{Introduction}
\setcounter{equation}{0}
\label{sec:introduction}
The dynamics of collapsing bodies is a subject which attracted
much attention amongst physicists since the very first formulation
of General Relativity.
However, the general problem of gravitational collapse is
extremely complicated and simplified models have therefore been
constructed which allow one to describe the dynamics at least
for the simplest configurations (for a review of analytical
solutions, see Ref.~\cite{adler}).
In this perspective, a lot of effort has been dedicated to the
study of the collapse of thin spherical shells
(see, e.g.~Refs.~\cite{grav,Israel,Pim}), for they
represent a quite general and simple model and can be regarded
as the starting point for the analysis of more complicated systems.
In recent years, attention has been focused on quantum
effects that can arise from strong gravitational fields
characterizing the final stages of the collapse and several
approaches have been attempted (for a very limited list, see,
e.g.~Refs.~\cite{Farhi,Balbi,Hajicek}).
\par
In particular, a formalism was introduced in Ref.~\cite{shellCQG}
which allows one to study the semiclassical behaviour of the shell by
adopting a ``minisuperspace'' approach~\cite{mini} to its dynamics
very much in the same spirit as that previously employed for the
Oppenheimer-Snyder model in Ref.~\cite{OS}.
Yet, many aspects of the collapse still need to be investigated
at the classical level, which may alter the quantum
(semiclassical) picture in a substantial way.
For example, the issue of whether the radial degree of freedom
of the shell can be described by a quantum wave function 
strictly depends on the fact that the matter composing
the shell be confined within a very small thickness
(i.e.~of the order of the \emph{Compton\/} wavelength of the
elementary particles composing the shell) around the
shell mean radius, so that the assumption of a coherent state
is reliable.
We refer to this issue as the ``stability'' problem for thick shells.
In fact, a realistic description of the collapse should involve
finite quantities (that is, a shell with finite thickness) whereas,
already for the classical dynamics, the very use of standard
junction equations~\cite{Israel} requires the validity of the
\emph{thin shell limit\/} (i.e.~the shell's thickness be
much smaller than its gravitational radius).
\par
In this paper we address the classical dynamics of a collapsing
spherical shell (which we call \emph{macroshell\/}) of proper mass $M$
with an inner massive core of mass $M_c$ and study the conditions
under which the matter composing the shell remains confined around
its mean radius.
In order to verify this property, we discretize the continuous distribution
of matter and consider the macroshell as if composed by a large
number $N$ of homogeneous nested spherical $\delta$-like
\emph{microshells\/}~\cite{shellPRD}.
Exploiting the spherical symmetry, we can adopt a Schwarzschild-like
coordinate system and the motion of the microshells can be described
by the usual ``areal'' radial coordinates $r_i$ ($i=1,\ldots,N$),
ordered so that $r_1<r_2<...r_N$.
Therefore, the thickness of the macroshell is given by $\delta=(r_N-r_1)$
and its mean radius is $R=(r_N+r_1)/2$.
Each microshell will be assumed to have the same
proper mass $m_i=m$ and a (possibly zero) electric charge $Q_i$.
We will limit our study outside the gravitational radius of the system,
$R_H=2\,G\,M_s/c^2$, where $M_s$ is the total ADM mass,
so that the radius of the innermost microshell $r_1$ will always
be greater than $R_H$.
\par
To summarise, the model will be constructed assuming the following
conditions:
\begin{enumerate}
\item
$M_c\gg M$.
The mass of the core is much bigger than the macroshell's,
so that the total mass-energy of the system is approximately
$M_s\simeq M_c$;
\label{1con}
\item
$m\ll Nm=M$.
Each microshell is much lighter than the macroshell;
\item
$(R>)\,r_1\gg \delta$.
The inner radius of the macroshell (and therefore the radius
of each microshell) is much larger than its thickness. 
This implies that the macroshell thickness also remains
much smaller than $R_H$ and this condition will limit
the temporal range of our analysis;
\item
$\displaystyle{\sum_{i=1}^{N}Q_i=0}$.
The macroshell is overall electrically neutral;
\item
$R(0)\gg R_H$ and
$\displaystyle{\left.\frac{\rmd R}{\rmd t}\right|_{t=0}\simeq 0}$.
The collapse starts in the weak field regime.
One can therefore take $t$ as the Schwarzschild time,
although it will then be more convenient to use the
(micro)shell proper times (for a detailed comparison,
see the Appendix~B in Ref.~\cite{shellPRD}).
\label{lcon}
\end{enumerate}
\par
As mentioned above, the classical equations of motion for a thin
shell in General Relativity can be obtained from the standard
Israel's junction equations.
In the present paper, we shall adopt a minisuperspace approach
in which the embedding metrics (interior and exterior with respect
to the shell) are held fixed and chosen to be particular solutions
of the (bulk) Einstein equations, and derive the shell's dynamical
equations from a variational principle applied to an effective
action.
This (equivalent) treatment will allow us to take into account the matter
internal degrees of freedom and can be extended to the semiclassical
level along the lines of Refs.~\cite{shellCQG,shellPRD}.
In fact, the equations thus obtained are equivalent to the junction
conditions, with the exception of one more equation which has no
counterpart as a junction condition and represents a secondary
constraint for the time variation of the gravitational mass of
the shell.
The relevant dynamical equation is a first order ordinary differential
equation for the shell radius which contains ADM parameters~\cite{ADM}
(those defining the inner and outer metrics).
In order to determine the shell trajectory, one can fix the initial shell
radius, but the initial velocity is then uniquely determined {\em if\/} the
ADM parameters were also fixed.
Alternatively (and more sensibly), one can set the initial radius and
velocity and adjust the ADM parameters correspondingly.
\par
The task of extending the above analysis to a system of (micro)shells
would be extremely involved in either cases.
If we choose to set the initial radii and velocities, then the ADM parameters
will have to be reset each time two shells cross.
Alternatively, if we set the ADM parameters once and for all, it is possible
that the system runs into some spurious (cusp or ``superficial'')
singularity when two shells cross~\cite{adler,singh}.
In light of a future extension to the semiclassical level, we shall
therefore find it more convenient to derive an effective Hamiltonian
for the motion of a single microshell in the mean gravitational
field generated by the core and all the other microshells treated as
a whole.
This will allow us to obtain a one-particle effective potential which
describes the tidal forces the microshell is subject to, and,
in particular, we will see that the binding potential varies with
time and confines the microshells within an ever decreasing
thickness (at least sufficiently away from the gravitational
radius of the system).
\par
We will work in units in which the speed of light $c=1$ but will
explicitly display the gravitational (Newton) constant $G$.
\section{Single shell dynamics}
We begin by reviewing the case of one $\delta$-like shell discontinuity.
The action for a generic matter-gravity system is given by~\cite{Haji}
\be
S=S_{EH}+S_{M}+S_{C}
\label{1.0}
\ ,
\ee
where $S_{EH}$ is the bulk Einstein-Hilbert action, $S_{M}$ the matter
action and $S_{C}$ accounts for boundary contributions.
\par
\begin{figure}[!t]
\begin{center}
\includegraphics[scale=0.7]{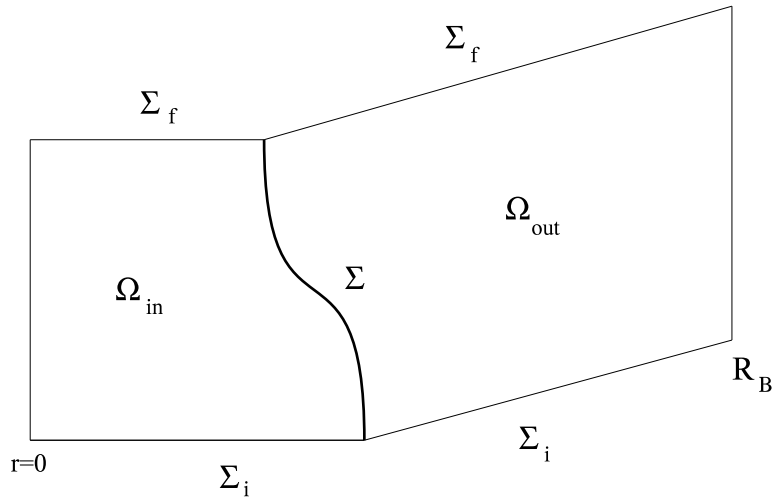}
\caption{A schematic view of the relevant space-time.
The surface at $r=R_B$ serves to bind the collapsing system inside
a ``box'' of finite spatial extension and the limit $R_B\to \infty$
can be eventually taken.
}
\label{s-t}
\end{center}
\end{figure}

In order to describe the motion of a shell, let us consider a space-time
volume $\Omega$ with boundary $\partial\Omega=\cup_k\Sigma_k$,
in which every $\Sigma_k$ represents a smooth three-dimensional submanifold
(see Fig.~\ref{s-t}).
The action takes the more explicit form (see Refs.~\cite{Farhi,Balbi}
and, for a thorough discussion, Ref.~\cite{Booth})
\be
\!\!\!\!
S&=&
\frac{1}{16\,\pi\,G}\,\int_{\Omega}
\rmd^4x\,\sqrt{-g}\,\mathcal{R}
+\int_{\Omega}\rmd^4x\,\mathcal{L}_m
\nonumber
\\
&&
+\frac{1}{8\,\pi\,G}\,\int_{\cup_k\Sigma_k}
\rmd^3x\,\sqrt{|^{(3)}g|}\,\mathcal{K}
+\frac{1}{8\,\pi\,G}\,\int_{\cap_k\Sigma_k}
\rmd^2x\,\sqrt{|^{(2)}g|}\,\sinh^{-1}(\eta)
\ ,
\label{1.1}
\ee
where $g$ is the determinant of the four-metric $g_{\mu\nu}$,
$\mathcal{R}$ the four-dimensional curvature scalar,
$\mathcal{K}$ the extrinsic curvature of the $3$-dimensional
hypersurface $\Sigma_k$ whose metric has determinant $^{(3)}g$
and $\eta$ is the scalar product of the unit normals to two
(non-smoothly) intersecting boundaries $\Sigma_k$ and $\Sigma_{k'}$
whose volume element is denoted by $|^{(2)}g|^{1/2}$.
Boundary terms have been included so as to obtain the equations
of motion by varying the metric inside $\Omega$ with vanishing
variations on the border $\partial\Omega$~\cite{grav,Booth}.
\par
In our case, we must also consider the three-dimensional
surface $\Sigma$ swept by the time-like sphere of area $4\,\pi\,R^2$
which splits the space-time $\Omega$ into two regions:
$\Omega_{\rm in}$, internal to the shell, and $\Omega_{\rm out}$,
external to the shell (where radiation can be included~\cite{action}).
Of course, there are an initial hypersurface $\Sigma_{\rm i}$
and a final hypersurface $\Sigma_{\rm f}$ with possible discontinuities at
the shell trajectory's end-points $\Sigma_{\rm i/f}\cap\Sigma$.
Adopting a minisuperspace approach, the metrics inside the
two regions $\Omega_{\rm in/out}$ are chosen to be spherically
symmetric solutions of the Einstein field equations and can be
written in the form
\be
\rmd s^2=-A\,\rmd t^2+A^{-1}\,\rmd r^2+r^2\,\rmd\Omega^2
\ ,
\label{1.2}
\ee
in which $r$ is the usual areal coordinate and $t$ is the time variable
$t_{\rm i}\le t\le t_{\rm f}$.
The most general solution of this kind (without external radiation)
is the Reissner-Nordstr\"om-(anti-)de~Sitter metric with
\be
A_{\rm in/out}=1-\frac{2\,G\,M_{\rm in/out}}{r}+\frac{Q_{\rm in/out}^2}{r^2}
\pm\frac{\Lambda_{\rm in/out}}{3}\,r^2
\ ,
\label{1.3}
\ee
where $M_{\rm in/out}$ is the ADM mass (proper mass plus gravitational energy)
enclosed in the sphere of radius $r$, $Q_{\rm in/out}$ represents the electric
charge inside the sphere of radius $r$, $\Lambda_{\rm in/out}$ the cosmological
constant (the signs before the last term account for both the de~Sitter
and the anti-de~Sitter metrics).
\par
In Ref.~\cite{action}, it was shown that, upon choosing a suitable ADM
foliation~\cite{ADM} of the space-time, the shell's trajectory can be expressed
as $r=R(\tau)$, where $\tau$ is an arbitrary time variable on the
shell's world volume $\Sigma$, and the three metric on
$\Sigma$ is given by
\be
\rmd s^2_{\Sigma}=-N^2\,\rmd\tau^2+R^2\,\rmd\Omega^2
\ ,
\label{1.3.1}
\ee
where $N=N(\tau)$ is the \emph{lapse function\/} of the shell.
On then integrating over space coordinates, the reduced effective action
(\ref{1.1}) for the shell can be written as
\be
S_{\mathrm{eff}}(N,R,M_{\rm out})&=&
\frac{1}{G}\,
\int_{\tau_{\rm i}}^{\tau_{\rm f}}\rmd\tau\,
\left[\beta\,R-\dot{R}\,R\tanh^{-1}\,
\left(\frac{\dot{R}}{\beta}\right)\right]_{\rm out}^{\rm in}
\nonumber
\\
&&
-\int_{\tau_{\rm i}}^{\tau_{\rm f}}\rmd\tau\,
\left[
4\pi\,N\,R^{2}\,\rho
-\frac{\dot{M}_{\rm out}R}{2\,A_{\rm out}}
+\frac{M_{\rm out}\dot{R}}{2A_{out}}
\right]
\nonumber
\\
&\equiv&
{\displaystyle \int_{\tau_{\rm i}}^{\tau_{\rm f}}
\rmd\tau\,L_{\mathrm{eff}}(N,R,M_{\rm out})}
\ ,
\label{1.4}
\ee
where $\beta\equiv\sqrt{A\,N^2+\dot{R}^2}$, the subscripts ``in''
and ``out'' now denote quantities calculated respectively
at $r=R-\varepsilon$ and $r=R+\varepsilon$ in the limit
$\varepsilon\to0^+$,
$\rho=\rho(R,M_{\rm out})$ is the mass-energy density of the shell,
and we have introduced the notation
$[F]^{\rm in}_{\rm out}\equiv F_{\rm in}-F_{\rm out}$.
\par
We point out that the action (\ref{1.4}) allows for the time variation
of the ADM mass $M_{\rm out}$ which is a canonical variable of the system,
and this corresponds to a radiating shell.
In this case, the region $\Omega_{\rm out}$ would be further divided into
two regions, one enclosing all the radiation and one ($\Omega_\infty$ in
Fig.~\ref{s-t}) devoid of both matter and radiation.
In the following Sections, the ADM mass calculated on the external surface
of the shell is constant and therefore the last two terms in the action
will be dynamically irrelevant.
However, we will give the more general expressions for the radiating case
below.
\par
Upon varying the action with respect to the canonical variables $N$,
$R$ and $M_{\rm out}$ one obtains the Euler-Lagrange equations of motion
for the shell in the form
\be
\frac{\delta L_{\mathrm{eff}}}{\delta \vec X}\equiv
\frac{\partial L_{\mathrm{eff}}}{\partial \vec X}
-\frac{\rmd}{\rmd\tau}
\left(\frac{\partial L_{\mathrm{eff}}}{\partial\dot{\vec X}}\right)
=0
\ ,
\label{1.5}
\ee
where $\vec X=(N,R,M_{\rm out})$ and
$\delta \vec X(\tau_{\rm i})=\delta \vec X(\tau_{\rm f})=0$.
Introducing the canonical momenta conjugated to the variables
$N$, $R$, $M_{\rm out}$ one obtains
\be
&&
P_{N}\equiv\frac{\partial L_{\mathrm{eff}}}{\partial \dot{N}}=0
\label{1.6}
\\
&&
P_{R}\equiv
\frac{\partial L_{\mathrm{eff}}}{\partial\dot{R}}=
-\left\{\frac{R}{G}\left[\tanh^{-1}\left(\frac{\dot{R}}{\beta}\right)
\right]_{\rm out}^{\rm in}
+\frac{M_{\rm out}}{2\,A_{\rm out}}\right\}
\label{1.7}
\\
&&
P_{M}\equiv
\frac{\partial L_{\mathrm{eff}}}{\partial\dot{M}_{\rm out}}
=\frac{R}{2\,A_{\rm out}}
\ .
\label{1.8}
\ee
We see from the first equation that $N$ is a Lagrange multiplier,
corresponding to the effective action being invariant under time
reparameterization on the shell.
Varying the effective action with respect to $N$ will
hence yield a primary constraint instead of a
real equation of motion.
Eqs.~(\ref{1.5}) thus read
\be
\frac{\delta L_{\mathrm{eff}}}{\delta N}
&=&
\frac{R}{G\,N}\left[\beta\right]_{\rm out}^{in}
-4\pi R^2\rho
=0
\label{1.9}
\\
\frac{\delta L_{\mathrm{eff}}}{\delta R}
&=&
\frac{1}{G}\left[\beta+\frac{1}{\beta}\left(R\ddot{R}
+\frac{N^{2}M}{R}-\frac{R\dot{N}\dot{R}}{N}
\right)\right]_{\rm out}^{\rm in}
+\frac{\dot{M}_{\rm out}}{\beta_{\rm out}}
\frac{\beta_{\rm out}-\dot{R}}{1-2\,M_{\rm out}/R}
\nonumber
\\
&&
-4\pi N\frac{\delta(R^2\rho)}{\delta R}
=0
\label{1.10}
\\
\frac{\delta L_{\mathrm{eff}}}{\delta M_{\rm out}}
&=&
4\,\pi\,N\,R^2\frac{\delta\rho}{\delta M_{\rm out}}
-\frac{\beta_{\rm out}-\dot{R}}{(1-2\,G\,M_{\rm out}/R)}=0
\ .
\label{1.11}
\ee
At this point, one can introduce the Hamiltonian of the system,
\be
H=P_{R}\,\dot{R}+P_{M}\,\dot{M}_{out}-L_{\mathrm{eff}}
=
4\,\pi\,N\,R^2\rho-\frac{R}{G}\left[\beta\right]_{\rm out}^{\rm in}
\ ,
\label{1.12}
\ee
and it becomes clear that Eq.~(\ref{1.9}) represents
the (secondary) Hamiltonian constraint in the form
\be
\frac{H}{N}=0
\ ,
\label{1.13}
\ee
whereas Eq.~(\ref{1.10}) can be written as the
time preservation of the constraint (\ref{1.13}),
\be
\frac{\rmd}{\rmd\tau}\left(\frac{H}{N}\right)=0
\ .
\label{1.14}
\ee
Making use of the invariance under time reparameterization
on the shell, it is convenient to choose the \emph{proper time\/}
gauge $N=1$, so that the dynamical evolution of the shell
is completely determined by the equation
\be
H=\frac{R}{G}\left[\beta\right]_{\rm out}^{\rm in}
-4\,\pi\,R^2\,\rho=0
\label{1.15}
\ee
and $\tau$ is the proper time on the shell.
If we consider a shell with constant ADM mass ${M}_{\rm out}$,
it seems obvious to identify the physical shell energy with
its proper mass and set $4\pi R^2\rho=M$.
With this substitution, Eq.~(\ref{1.15}) reads
\be
M=\frac{R}{G}\,\left(\sqrt{A_{\rm in}+\dot{R}^{2}}
-\sqrt{A_{\rm out}+\dot{R}^{2}}\right)
\ ,
\label{1.16}
\ee
where we have used the explicit expression for $\beta$ and
$\dot{R}=\rmd R/\rmd\tau$ denotes the total derivative with respect
to the shell's proper time $\tau$.
\subsection{Dynamical equation and initial conditions}
\label{in_cond}
\begin{figure}[!t]
\begin{center}
\raisebox{5cm}{$|\dot{R}|$}
\includegraphics[scale=0.7]{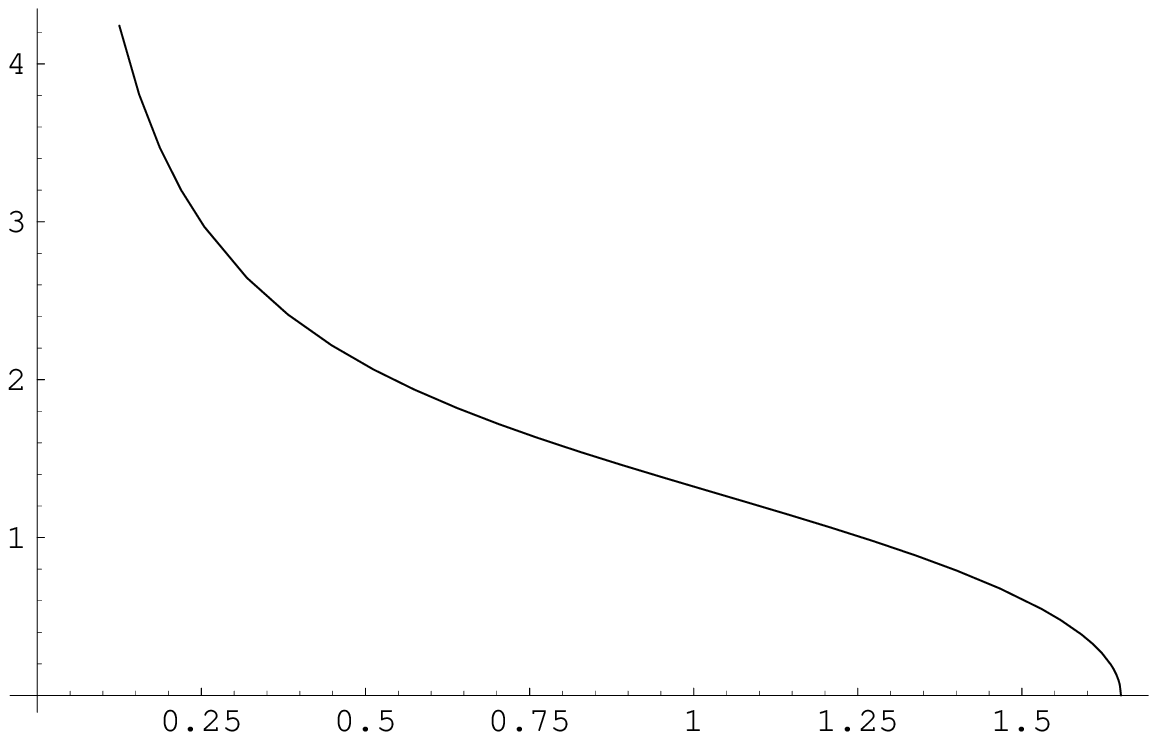}
\\
\hspace{8cm}$R/M$
\caption{Evolution of the macroshell's velocity with the radius
for $M_s=M$, $M_c=0$ and $\Lambda=1$ (in units with $G=1$).
}
\label{RdR}
\end{center}
\end{figure}
On squaring twice the above Eq.~(\ref{1.16}) and rearranging we have
\be
\dot R^{2}
=\frac{R^{2}}{4\,M^{2}\,G^{2}}\,
\left(A_{\rm in}-A_{\rm out}\right)^{2}
-\frac{1}{2}\,
\left(A_{\rm in}+A_{\rm out}\right)
+\frac{M^{2}\,G^{2}}{4\,R^{2}}
\ .
\label{1.17}
\ee
It is then clear that the shell's trajectory will be uniquely specified by the
initial condition $R(\tau=0)=R_0$ if the metric functions $A_{\rm in/out}$
have fixed parameters.
One is therefore given two options:
\begin{description}
\item[a)]
fix the metric parameters and the initial radius $R(0)=R_0$;
\item[b)]
set the initial radius $R(0)=R_0$ and velocity $\dot R(0)=\dot R_0$,
and adjust the metric parameters accordingly.
\end{description}
\par
For example, let us consider a macroshell collapsing in a
Schwarzschild-de~Sitter background, with metrics
\be
&&
A_{\rm in}=
1-\frac{2\,G\,M_c}{r}
+\frac{\Lambda}{3}\,r^2
\nonumber
\\
\label{1.17.2}
\\
&&
A_{\rm out}=
1-\frac{2\,G\,M_{s}}{r}+\frac{\Lambda}{3}\,r^2
\ ,
\nonumber
\ee
where $M_c$ is the ADM mass of the core and
$M_s$ the total ADM mass of the system~\footnote{Note that
$M_s-M_c>0$ is not equal to the shell's proper mass $M$
in general.
In fact, it can be smaller than $M$ for bound orbits since
it also contains the shell's negative gravitational
``potential energy''.}.
Substituting Eqs.~(\ref{1.17.2}) into Eq.~(\ref{1.17}),
we obtain
\be
\dot R^{2}
=\left(\frac{M_s-M_c}{M}\right)^2
-1+\frac{G\,\left(M_s+M_c\right)}{R}
+\frac{G^2\,\,M^2}{4\,R^2}-\frac{\Lambda}{3}\,R^2
\ ,
\label{1.17.4}
\ee
which gives the time evolution of the shell's velocity in terms
of its radius $\dot{R}\left(R(\tau)\right)$.
Clearly, (as Fig.~\ref{RdR} also shows) that equation describes
the collapse of a shell starting from its maximum radius
$R_0=R_{\rm max}(M,M_s,M_c,\Lambda)$ with zero initial velocity
$\dot{R}_0=0$.
In order to be able to set the initial velocity independently of the initial
radius, one can relax one of the metric paramaters, e.g.,~the total ADM mass
$M_s$, which will therefore be determined as $M_s=M_s(R_0,\dot R_0;\Lambda,M,M_c)$
by Eq.~(\ref{1.17.4}) at $\tau=0$.
\subsection{Generalised dynamics}
\label{gen_dyn}
As previously noted, one cannot set the initial radius and velocity
if the metric paramaters have already been fixed.
This, for example, rules out the possibility of setting $M_s=M_c+M$,
$\Lambda=0$ and have the shell start collapsing with zero velocity at
finite radius.
\par
Another possibility is to consider the presence of extra
(non-gravitational) forces which certainly come into play in the
description of astrophysical objects, and allow for equilibrium
configurations of the collapsing matter at finite radius (for example,
the radiation pressure for main-sequence stars and the degeneracy
pressure of the electrons and, at later stages, of the neutrons
for compact stars).
A way to model that is to introduce an {\em ad hoc\/} force which acts
upon the system {\em before\/} the collapse begins and manages to keep
the radius $R$ fixed (for $\tau<0$). 
Such a force will be switched off at $\tau=0$ and the shell then
collapses under the pull of just the gravitational force.
One can study this (generalized) dynamics of the shell by
adding a constant $V_0$ to the shell effective Hamiltonian,
\be
\!\!\!\!\!\!\!\!\!\!\!\!\!\!\!\!\!\!\!\!\!\!\!\!\!
H_{\mathrm{eff}}
&=&
\frac{M}{2}\,\dot R^{2}+V(R)+V_0
\nonumber
\\
\!\!\!\!\!\!\!\!\!\!\!\!\!\!\!\!\!\!\!\!\!\!\!\!\!
&=&
\frac{M}{2}\,
\left[\dot R^{2}
-\left(\frac{M_s-M_c}{M}\right)^2
+1
-\frac{G\,\left(M_s+M_c\right)}{R}-\frac{G^2\,M^2}{4\,R^2}
+\frac{\Lambda}{3}\,R^2+\tilde f_0
\right]
\ ,
\label{1.17.8}
\ee
where 
\be
\tilde f_0&=&
\frac{2}{M}\,\left[V(R_0)-V_0\right]
\nonumber
\\
&=&
\left(\frac{M_s-M_c}{M}\right)^2
-1
-\frac{G\,\left(M_s+M_c\right)}{R_0}
+\frac{G^2\,M^2}{4\,R_0^2}
-\frac{\Lambda}{3}\,R_0^2
\ ,
\ee
is now an arbitrary constant.
We can now set the initial conditions $R(0)= R_0$ and
$\dot{R}(0)= 0$ for a given $R_0$.
Let us finally remark that the Hamiltonian constraint is still
given by
\be
H_{\mathrm{eff}}=0
\ ,
\label{1.17.10}
\ee
and the analysis then proceeds straightforwardly. 
In fact, Eq.~(\ref{1.17.8}) for a non-radiating shell is still compatible with
the constraint~(\ref{1.14}), since $\tilde f_0$ and all the metric parameters
are constant in such a case and the dynamical system remains well defined.
\section{The macroshell model}
\setcounter{equation}{0}
In our model, the macroshell is collapsing in a black-hole
background and the total ADM mass of the system is constant.
In order to study the stability of the matter composing the shell,
its internal structure must be taken into account and therefore
the macroshell is viewed as if composed by many nested microshells.
We expect the matter distribution inside the macroshell to alter
locally the black-hole background and we study the motion of a
single microshell in the mean gravitational field generated by
the remaining $(N-1)$ microshells.
\par
\begin{figure}[!t]
\begin{center}
\includegraphics[scale=0.5]{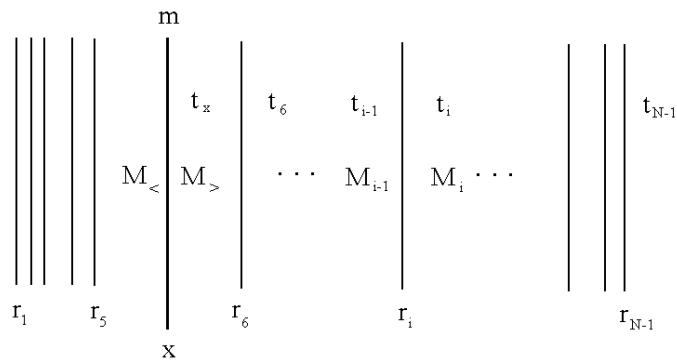}
\caption{Internal structure of the macroshell.
We have denoted with $M_<\equiv M_{\rm in}$ and $M_>\equiv M_{\rm out}$
the ADM mass calculated respectively on the internal and on the
external surface of the microshell of radial coordinate $x$.
\label{macro}}
\end{center}
\end{figure}
In order to do this, let us single out a microshell of radius $x$
and proper mass $m=M/N$ (see Fig~\ref{macro}).
When $x\ne r_i$, we can make use of Eq.~(\ref{1.15}) with
$4\,\pi\, x^2\,\rho=m$, substitute the expression for
$\beta$, square twice and rearrange so as to obtain
\be
\dot x^{2}
=\frac{x^{2}}{4\,m^{2}\,G^{2}}\,\left(A_{<}-A_{>}\right)^{2}
-\frac{1}{2}\,\left(A_{<}+A_{>}\right)
+\frac{m^{2}\,G^{2}}{4\,x^{2}}
\ ,
\label{1.17.1}
\ee
where $\dot{x}=\rmd x/\rmd\tau$ denotes the total derivative with
respect to the shell's proper time $\tau$.
Then, on multiplying Eq.~(\ref{1.17}) by $m/2$,
we obtain the equation of motion for a microshell in the form
of an effective Hamiltonian constraint, induced by the canonical
Hamiltonian constraint~(\ref{1.15}), namely
\be
H_{\mathrm{eff}}
=\frac{m}{2}\,\dot x^{2}+V_{\mathrm{eff}}(x)
=0
\ ,
\label{1.18}
\ee
and comparison with Eq.~(\ref{1.17}) immediately yields
\be
V_{\mathrm{eff}}(x)
=\frac{m}{4}\,\left(A_{<}+A_{>}\right)
-\frac{x^{2}}{8\,m\,G^{2}}\,\left(A_{<}-A_{>}\right)^{2}
-\frac{m^{3}\,G^{2}}{8\,x^{2}}
\ .
\label{1.19}
\ee
In order to proceed, it would now be necessary to substitute the explicit
expressions for $A_{<}$ and $A_{>}$, which requires the knowledge of
the microshells distribution inside the macroshell.
As we discussed in Section~\ref{in_cond}, there are two equivalent
ways of determining the evolution of $x=x(\tau)$, one which allows one
to set both $x(0)=x_0$ and $\dot x(0)=\dot x_0$ (with metric parameters
correspondingly determined), and one in which the metric parameters are
fixed and just $x(0)=x_0$ can be set.
However, the former option is presently further complicated by the presence
of more than one shell.
In fact, whenever two shells cross (i.e.~at the time $\tau=\tau_c$
when $x(\tau_c)=r_i$ for some $i=1,\ldots,N-1$),
the metric parameters must be re-evaluated using $x(\tau_c)$ and
$\dot x(\tau_c)$ as new initial conditions~\footnote{We do not intend
to discuss the issue of shell crossings in any detail here.
What happens to two shells when they collide would in fact require
the implementation of specific matter properties (such as the detailed
treatment of non-gravitational interactions).}.
Since $\tau_c$ in turn depends on the trajectory before the crossing,
this also means that the system keeps memory of its entire evolution.
For this reason, and because we want to be able to extend the treatment
to the semiclassical level in the future, we find it more convenient
to pursue the other option and fix the metric parameters once and for all.
This may imply that the generalised dynamics of Section~\ref{gen_dyn}
will be used in order to allow us to impose arbitrary initial velocity
(e.g.~in Section~\ref{stability}).
We shall however not display the arbitrary constant $\tilde f_0$ in
order to keep the expressions simpler.
\par
We then choose the metric parameters according to the assumptions
(\ref{1con})-(\ref{lcon})~as outlined in the Introduction.
First of all, since $m\ll M\ll M_s$, we shall take
$M_{>}-M_{<}\simeq m$, which amounts to $\dot x(0)\simeq 0$
and is consistent with the choice of having the microshells
confined within a thickness $\delta$.
With this assumption both $M_{<}$ and $M_{>}$ become functions
of the microshells distribution $\{(r_i,M_i),\ i=1,\ldots,N\}$,
but, since it is unknown {\em a~priori\/}, the evaluation of the
potential remains involved.
A reliable approximation is therefore to adopt a mean field approach
and study the motion of the microshell when $x<r_1$ or $x>r_1+\delta$;
in the thin shell limit for the macroshell, all the remaining $(N-1)$
microshells can be viewed as forming a shell of proper mass $(M-m)$ and,
if we denote its average radial coordinate by $X$, we can obtain an
equation of motion analogous to Eq.~(\ref{1.17}).
In agreement with our gauge choice,
we will use the proper times of the (two) shells to foliate the
space-time and therefore consider the dynamical variables $x$ and $X$
as functions of the same variable $\tau$.
We remark that, if the microshell bears an electric charge $Q$,
then the shell of mass $(M-m)$ must have a charge $-Q$ so that
the total macroshell of mass $M$ has zero charge.
\par
\begin{figure}[!t]
\begin{center}
\includegraphics[scale=0.5]{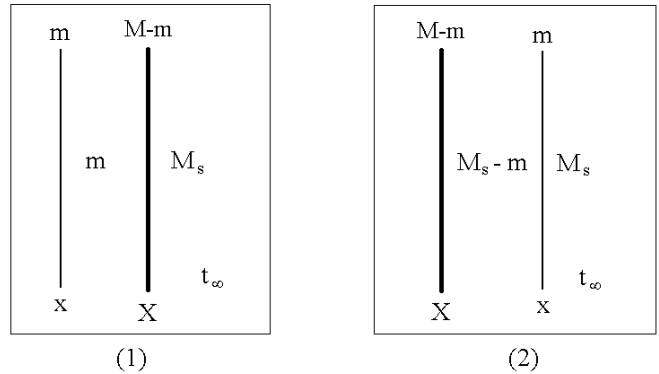}
\caption{Equivalent configuration in the \emph{thin-shell\/} limit:
(1) $x<X$; (2) $x>X$.
\label{ext}}
\end{center}
\end{figure}
One has therefore two distinct situations for $x<X$ and $x>X$
(see cases (1) and (2) in Fig.~\ref{ext}) and two distinct
effective potentials:
\par
For $x<X$, the region $R_H<r<x$ has a Schwarzschild-de~Sitter
metric with ADM mass $(M_s-M)$, the region $x<r<X$ has a
Reissner-Nordstr\"om-de~Sitter metric with ADM mass $(M_s-M+m)$
and electric charge $Q$, while the outer region $r>X$ has again
Schwarzschild-de~Sitter metric with ADM mass $M_s$.
We therefore get 
\be
\!\!\!\!\!\!\!\!
V_{\mathrm{eff}}^x(x)
&=&
\frac{m}{2}\,\left[-\frac{G}{x}\,\left(2\,M_s-2\,M+m\right)
+\frac{Q^2}{G\,m\,x}-\frac{G^2\,m^2}{4\,x^2}
\right.
\nonumber
\\
&&
\left.
\phantom{\frac{m}{2}\ \,}
+\frac{Q^2}{2\,x^2}\left(1-\frac{Q^2}{2\,G^2\,m^2}\right)
+\frac{\Lambda}{3}\,x^2\right]
\label{1.20}
\ee
\be
\!\!\!\!\!\!\!\!
V_{\mathrm{eff}}^X(X)&=&
\frac{M-m}{2}\,\left[
-\frac{G}{X}\,\left(2\,M_s-M+m\right)
-\frac{Q^2}{G\,(M-m)\,X}-\frac{G^2\,(M-m)^2}{4\,X^2}
\right.
\nonumber
\\
&&
\left.
\phantom{\frac{M-m}{2}\ \,}
+\frac{Q^2}{2\,X^2}\,\left(1-\frac{Q^2}{2\,G^2\,(M-m)^2}\right)
+\frac{\Lambda}{3}\,X^2\right]
\ ,
\label{1.21}
\ee
where Eq.~(\ref{1.20}) represents the effective potential
for the microshell of proper mass $m$ and radius $x$,
while Eq.~(\ref{1.21}) is the effective potential for the
macroshell of proper mass $(M-m)$ and radius $X$.
\par
The same procedure can be followed for $x>X$.
In this case the region $r<X$ has a Schwarzschild-de~Sitter
metric with ADM mass $(M_s-M)$, the region $X<r<x$ has a
Reissner-Nordstr\"om-de~Sitter metric with ADM mass $(M_s-m)$
and electric charge $-Q$, while the outer region $r>x$ has
Schwarzschild-de~Sitter metric with ADM mass $M_s$.
This yields
\be
\!\!\!\!\!\!\!\!
V_{\mathrm{eff}}^x(x)
&=&
\frac{m}{2}\,\left[-\frac{G}{x}\,\left(2\,M_s-m\right)
-\frac{Q^2}{G\,m\,x}-\frac{G^2\,m^2}{4\,x^2}
\right.
\nonumber
\\
&&
\left.
\phantom{\frac{m}{2}\ \,}
+\frac{Q^2}{2\,x^2}\,\left(1-\frac{Q^2}{2\,G^2\,m^2}\right)
+\frac{\Lambda}{3}\,x^2\right]
\label{1.22}
\ee
\be
\!\!\!\!\!\!\!\!
V_{\mathrm{eff}}^X(X)
&=&
\frac{M-m}{2}\,\left[-\frac{G}{X}\,\left(2\,M_s-M-m\right)
+\frac{Q^2}{G\,(M-m)\,X}-\frac{G^2\,(M-m)^2}{4\,X^2}
\right.
\nonumber
\\
&&
\left.
\phantom{\frac{M-m}{2}\ \,}
+\frac{Q^2}{2\,X^2}\,\left(1-\frac{Q^2}{2\,G^2\,(M-m)^2}\right)
+\frac{\Lambda}{3}\,X^2\right]
\ .
\label{1.23}
\ee
\par
Since Eq.~(\ref{1.18}) must hold for both the microshell
and the macroshell, for each one of the two cases presented
above we can write a total effective Hamiltonian for the system
which satisfies the constraint
\be
H_{\mathrm{eff}}^{\rm tot}(x,X)&\equiv&
H_{\mathrm{eff}}^{x}(x)+H_{\mathrm{eff}}^{X}(X)
\nonumber
\\
&\equiv&
\frac{m}{2}\,\dot x^2
+\frac{M-m}{2}\,\dot X^2
+V_{\mathrm{eff}}^x(x)+V_{\mathrm{eff}}^X(X)
\nonumber
\\
&\equiv&
\frac{m}{2}\,\dot x^2
+\frac{M-m}{2}\,\dot X^2
+V_{\mathrm{eff}}^{\rm tot}(x,X)=0
\ .
\label{1.24}
\ee
\par
The dynamics has now been reduced to that of a two-body system,
so that we can separate the motion of the center of mass from
the relative motion by introducing the
\emph{center of mass\/} and \emph{relative\/}
coordinates given by
\be
\left\{ {\displaystyle \begin{array}{l}
\bar{r}\equiv x-X
\\
\\
{\displaystyle R\equiv\frac{m}{M}\,x
+\frac{\mu}{m}\,X}
\ ,
\end{array}}\right.
\label{1.25}
\ee
where $\mu\equiv {m\,\left(M-m\right)}/{M}$ is the
\emph{reduced mass\/} of the system.
The Hamiltonian constraint~(\ref{1.24}) then takes the form
\be
H^{(\tau)}(\bar{r},R)
\equiv
\frac{\mu}{2}\,\dot{\bar r}^2
+\frac{M}{2}\,\dot R^2+
V^{(\tau)}(\bar{r},R)=0
\ ,
\label{1.26}
\ee
where the superscript $\tau$ is to recall the time dependence
of the Hamiltonian through the variables $\bar{r}$ and $R$.
\par
Since one component of the system (the macroshell) is much
more massive than the other one (the microshell) we have,
to lowest order in $m/M$, $\mu\simeq m$ and we may thus
consider the motion of the microshell in the center of mass
system to be described by the relative coordinate $\bar{r}$
in much the same way as one treats the motion of the electron
in the hydrogen atom.
Indeed, since the macroshell is freely falling in the black-hole
background, its (mean) radius $R$ and proper time $\tau$
define a locally inertial frame.
We would thus like to subtract the center of mass motion
and obtain a one-particle Hamiltonian for the microshell.
In so doing, we obtain an effective Hamiltonian for the whole
macroshell of proper mass $M$ and radius $R$ by making use of
Eqs.~(\ref{1.18}) and (\ref{1.19}).
In this case we just have two regions, both with Schwarzschild-de~Sitter
metric,
$r<R$ with ADM mass $(M_s-M)$ and $r>R$ with ADM mass $M_s$,
\be
H_M^{(\tau)}(R)
&=&\frac{M}{2}\,\dot R^2+V_M^{(\tau)}(R)
\nonumber
\\
&=&
\frac{M}{2}\,\left[\dot R^2
-\frac{2\,G}{R}\,\left(M_s-\frac{M}{2}\right)
-\frac{M^2\,G^2}{4\,R^2}+\frac{\Lambda}{3}\,R^2\right]
\ .
\label{1.27}
\ee
Finally, we obtain the one-particle Hamiltonian constraint
for the microshell as
\be
H^{(\tau)}_m(\bar{r})\equiv H^{(\tau)}(\bar{r},R)-H_M^{(\tau)}(R)
\simeq
\frac{m}{2}\,\dot{\bar r}^2
+V_{m}^{(\tau)}(\bar{r})=0
\ ,
\label{1.28}
\ee
where the potential $V_{m}^{(\tau)}$ may contain a term
$\tilde f_0$ so as to ensure that $\bar r(0)\simeq \delta/2$
{\em and\/} $\dot {\bar r}(0)\simeq 0$,
according to the discussion of the generalised dynamics
in Section~\ref{gen_dyn}.
Further, the variable $R$ in the potential $V_{m}^{(\tau)}$
is now to be considered as a (time-dependent) parameter.
We note here that this approximation is justified if the typical
times of evolution of the macroshell radius $R$
are much larger than the period of a typical oscillation
of the microshell around $R$.
\par
The explicit expression for the potential
$V_{m}^{(\tau)}(\bar{r})$ is extremely complicated and we
deem pointless to write it explicitly.
Instead, in the next Section, we will plot its behaviour in
situations of particular interest.
\section{Shell stability}
\label{stability}
\setcounter{equation}{0}
In order to study the stability of the shell we suppose that
for $\tau=0$ all the matter is localized within a thickness
$\delta\ll R_H\ll R$ and the shell has negligible (initial)
velocity $\dot R(0)\simeq0$.
This corresponds to a microshell confined in the region
$\left|\bar{r}\right|<\delta$ with maximum kinetic energy
\be
\frac{m}{2}\,\dot{r}^2_{\rm max}
\simeq
V_{m}^{(\tau)}(\bar r=\pm\delta)-V_{m}^{(\tau)}(\bar r=0)
\ ,
\label{1.29}
\ee
provided the value $V_{m}^{(\tau)}(\pm\delta)$ is less than
the (possible) maximum of $V_{m}^{(\tau)}$.
\par
\begin{figure}[!t]
\begin{center}
\includegraphics[scale=0.7]{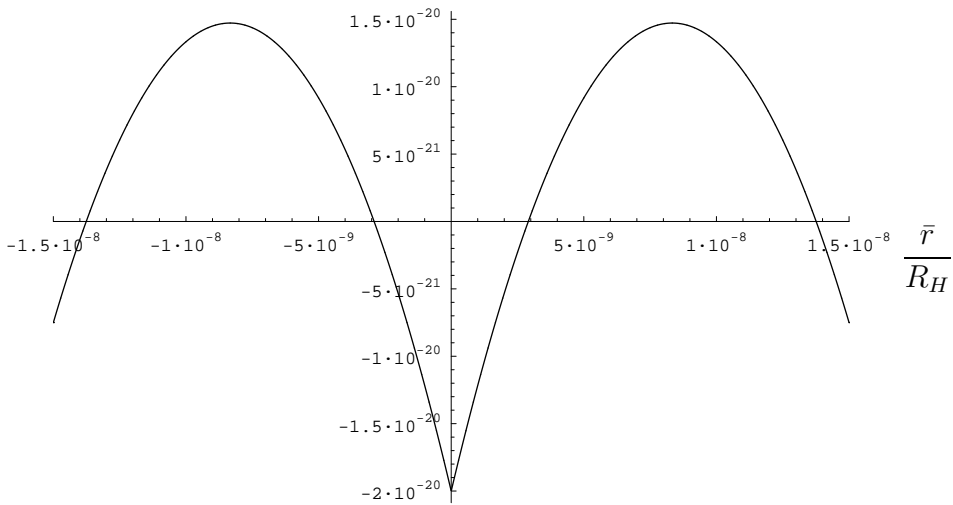}
\raisebox{3.5cm}{$\displaystyle\frac{\bar r}{R_H}$}
\caption{Effective potential $V_m^{(0)}(\bar r)/m$
for a macroshell thickness $\delta=6\cdot 10^{-9}\,R_H$
and macroshell radius $R=10\,R_H$.
The metric parameters are
$M_c=3\,M_\odot$,
$M\simeq 8.5\cdot 10^{-9}\,M_c
\simeq2.5\cdot 10^{-8}\,M_\odot$,
$m\simeq 4\cdot 10^{-9}\,M\simeq 1\cdot 10^{-16}\,M_\odot$,
$M_s=M_c+M$, and $\Lambda=Q=0$,
($M_\odot\simeq 2\cdot 10^{33}\,$g is the solar mass
and $R_H\simeq 9\cdot 10^{4}\,$cm).
\label{Vsample}}
\end{center}
\end{figure}
In Fig.~\ref{Vsample}, we plot the effective potential for a
typical case (see the caption for the values of the parameters) at
the (relatively) large value of $R=10\,R_H$ and macroshell thickness
$\delta\simeq 6\cdot 10^{-9}\,R_H$ (we have chosen arbitrary constants
so that $V_m^{(0)}(\pm\delta)=0$).
From the graph, it appears that the microshell is classically bound
around the ``centre of mass'' of the macroshell, in agreement with
previous results to first order in the small $|\bar r|/R$
expansion~\cite{shellPRD}.
This means that if the microshell is initially placed close enough
to the macroshell, the mutual common gravitational attraction will
keep them together.
However, for $|\bar r|\gg \delta$, the potential bends down and
starts decreasing.
This effect cannot be seen in a perturbative expansion for 
small $|\bar r|/R$ but is encoded in the exact potential.
One can understand such an effect on considering that if the
microshell started with a radius sufficiently larger or smaller
than the macroshell's, the mutual attraction would not be sufficient
to overcome the difference in collapsing velocities and their
``distance'' $|\bar r|$ would increase along the collapse.
We can note in passing that, as a consequence of the form of the
effective potential, one expects that the quantum-mechanical
probability for the microshell to tunnel out the well
(i.e.,~from $|\bar r|<\delta$ to $|\bar r|\gg \delta$)
is not zero and (approximately) symmetrical for the
cases of a microshell falling inwards or outwards.
Strictly speaking, bound states of microshells would therefore
be just quantum-mechanically {\em meta\/}-stable.
\par
\begin{figure}[!t]
\begin{center}
\includegraphics[scale=0.6]{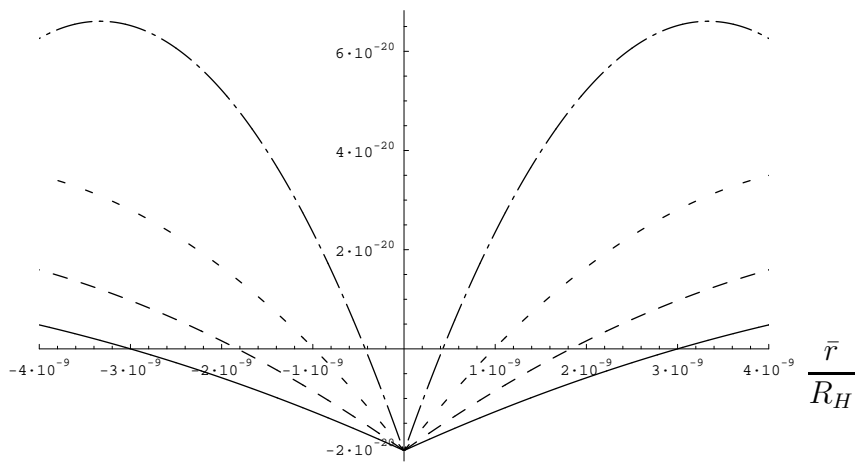}
\raisebox{1.3cm}{$\displaystyle\frac{\bar r}{R_H}$}
\caption{Effective potential $V_{m}^{(\tau)}/m$
for the same parameters as in Fig.~\ref{Vsample} and
$R=10\,R_H$ (solid line),
$R=8\,R_H$ (dashed line),
$R=6\,R_H$ (dotted line), and
$R=4\,R_H$ (dash-dotted line).
\label{noQL}}
\end{center}
\end{figure}
During the collapse, the velocity of the macroshell $\dot{R}$
increases while its radius $R$ shrinks and the effective potential
for the microshell thus changes.
This behaviour is easily understood as the increasing of
tidal forces between different parts of the macroshell.
One might therefore wonder how the time evolution of the binding
potential affects the macroshell's thickness and if, eventually,
the microshells can escape thus spreading out the system.
From the plots in Fig.~\ref{noQL}, we can see that the potential
$V_{m}^{(\tau)}$ remains symmetric around the centre but appears
to become steeper and steeper as the shell approaches
the gravitational radius.
In fact, to a given maximum kinetic energy~(\ref{1.29}), there
correspond smaller and smaller values of $\delta$ for decreasing
values of $R$ (moreover, it is easy to see that $\delta/R$
actually decreases, since $\delta$ decreases faster than $R$).
On the other hand, one may also note that the ``width'' of the forbidden
region decreases (whereas the maxima of the potential increase)
and it is thus possible that, at the quantum level,
the meta-stable states have different lifetimes.
A definite conclusion on this point would need a more complete
semiclassical analysis beyond the scope of the present work. 
\par
\begin{figure}[!ht]
\begin{center}
\includegraphics[scale=0.6]{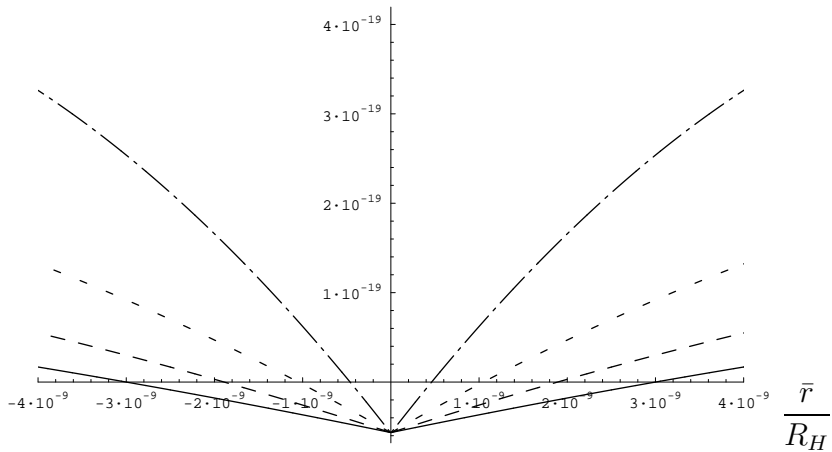}
\raisebox{0.5cm}{$\displaystyle\frac{\bar r}{R_H}$}
\caption{Effective potential $V_{m}^{(\tau)}/m$ for the same
parameters as in Fig.~\ref{Vsample} except that $Q=10^{29}\,e$
($e$ being the electron's charge) and
$R=10\,R_H$ (solid line),
$R=8\,R_H$ (dashed line),
$R=6\,R_H$ (dotted line), and
$R=4\,R_H$ (dash-dotted line).
\label{Q}}
\end{center}
\end{figure}
\begin{figure}[!t]
\begin{center}
\includegraphics[scale=0.6]{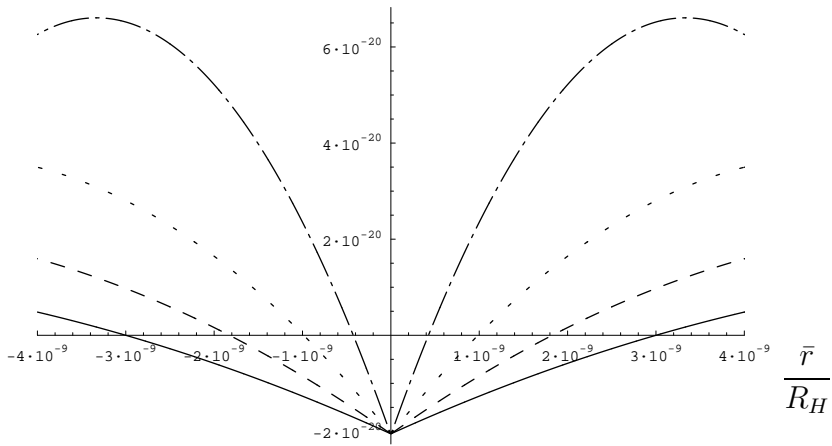}
\raisebox{1cm}{$\displaystyle\frac{\bar r}{R_H}$}
\caption{Effective potential $V_{m}^{(\tau)}/m$ for the same
parameters as in Fig.~\ref{Vsample} except that
$\Lambda=10^{-56}\,$cm$^{-2}$ and
$R=10\,R_H$ (solid line),
$R=8\,R_H$ (dashed line),
$R=6\,R_H$ (dotted line), and
$R=4\,R_H$ (dash-dotted line).
\label{L}}
\end{center}
\end{figure}
For completeness, we plot the evolving potential
when opposite electric charges are carried by the microshell
and macroshell in Fig.~\ref{Q},
and when a cosmological constant is included in Fig.~\ref{L}.
\section{Conclusions}
\setcounter{equation}{0}
\label{conc}
We have examined the collapse of a macroscopic self-gravitating
spherical shell of matter in the presence of a massive core in
the framework of General Relativity.
The (macro)shell is assumed to be formed by a large number of thin
(micro)shells initially bundled within a very short thickness.
The system is then shown to remain classically stable, since the effective
potential which drives the microshells becomes more and more binding
as the macroshell approaches the gravitational radius of the system
(but remains sufficiently away from it so that our approximations
$R-R_H\gg\delta$ and $|\dot R|<1$ hold).
Further, including a (reasonably valued) cosmological constant does
not alter this behaviour, nor does the presence of electric charges
carried by the microshells.
\par 
A semiclassical analysis may however change the above conclusion,
since the bound states of microshells are just meta-stable and there
may be significant changes in their lifetimes (or other, non-adiabatic,
effects) as the macroshell collapses.
In Refs.~\cite{shellPRD,ac}, it was shown that the microshells can
indeed be described by means of localized quantum mechanical wave
functions, however in the linear approximation (first order in
$|\bar r|/R$) for the effective potential.
We have now found that, for sufficiently large values of $|\bar r|$
(but still much smaller than $R$),
the complete effective potential reaches maxima and then decreases,
thus suggesting a non-vanishing probability for the microshells to
tunnel out. 
It was also found in Refs.~\cite{shellPRD,ac} that, upon coupling
the microshell wave functions to a radiation field, the shell
spontaneously emit during the collapse as a non-adiabatic effect.
A further analysis of such semiclassical effect in light of the
present results is therefore in order, which could also affect
the usual thermodynamical description~\cite{thermo}.
\par
Let us conclude by mentioning that inclusion of (or a possible
relation with) the Hawking radiation~\cite{hawk} as the shell
approaches the gravitational radius still remains to be investigated.
\section*{References}
\end{document}